\documentclass[prb,aps,showpacs,preprint]{revtex4}
\usepackage{epsfig}
\usepackage{amssymb}
\usepackage{amsmath}
\usepackage{graphicx}
\usepackage{amsfonts}
\usepackage[normalem]{ulem}

\begin{document}
\title{Coexistence of Superconductor and Spin-Density Wave in (TMTSF)$_2$ClO$_4$: Spatial Structure of the Two-phase State}
\author{Ya.\,A.~Gerasimenko$^1$}
\author{S.\,V.~Sanduleanu$^{1,2}$}
\author{V.\,A.~Prudkoglyad$^1$}
\author{A.\,V.~Kornilov$^1$}
\author{J.~Yamada$^3$}
\author{J.\,S.~Qualls$^4$}
\author{V.\,M.~Pudalov$^{1,2}$}
\affiliation{$^1$P.\,N. Lebedev Physical Institute of the RAS, Moscow, 119991, Russia}
\affiliation{$^2$Moscow Institute of Physics and Technology, Dolgoprudny, 141700, Russia}
\affiliation{$^3$University of Hyogo, Hyogo, 678-1297, Japan}
\affiliation{$^4$Sonoma State University,  Rohnert Park, CA 94928, USA}
\begin{abstract}
We report comprehensive (magneto)transport studies of the two-phase state in (TMTSF)$_2$ClO$_4$, where superconducting (SC) phase coexists with spin-density wave insulator (SDW). By tuning the degree of ClO$_4$ anion ordering in controlled manner we smoothly suppress the SDW state and study resulting evolution of the SC phase spatial texture. We find that as SDW is suppressed, SC regions initially appear inside the SDW insulator in a form of filaments extended in the interlayer direction and further merge into the two-dimensional sheets across the most conducting axis of the crystal. We demonstrate that almost all our results can be explained within the soliton phase model, though with several assumptions they can also be related with the creation of non-uniform deformations. We believe that the anisotropy is intrinsic to SC/SDW coexistence in various quasi one-dimensional superconductors.
\end{abstract}


\date{\today}

\maketitle

\section{Introduction}
Many low-dimensional and layered unconventional superconductors have similar phase diagrams in the sense that the superconductor phase neighbors the insulating (often magnetically ordered)  phase, and the two phases may coexist with each other in a finite interval of doping or pressure \cite{Lebed08, Stewart11, Johnston10, Chubukov12}. The well-known example is the  quasi one-dimensional (Q1D) organic (TMTSF)$_2$PF$_6$\cite{Lebed08} compound (hereafter denoted as PF6), where  superconductivity (SC) coexists with antiferromagnetically ordered spin-density wave (SDW) state in a narrow pressure region on the $T-P$ phase diagram.
This interesting region was extensively studied using macroscopic (transport\cite{Vuletic02,Kornilov04,Lee05}) and local (NMR\cite{Yu02,Lee05}) probes. Both types of measurements provide solid evidence for segregation of SC and SDW phases in real space. The phase segregation at the first glance can be explained within the framework of simple thermodynamic arguments, as a result of interplay between elastic and SDW energies \cite{Vuletic02,Kornilov04,Lee05} at the first order transition. This is in agreement with the transport measurements suggesting a large macroscopic spatial size of the metal(SC) regions \cite{Vuletic02,Lee05}. NMR measurements however show deviations from this simple picture \cite{Yu02,Lee05}. An order of magnitude increase of the SC critical field in the two-phase state, as determined from transport measurements \cite{Lee02}, also suggests a smaller size of the SC domains. The above inconsistency of the experimental data suggests that the spatial structure of the two phase state may be highly anisotropic and non-trivial, and therefore requires more targeted studies.

A number of theories was put forward to explain the coexistence of metal(SC) and SDW phases in PF6. However, some of them do not envisage any specific spatial texture. Among them is the SO(4) symmetry treatment \cite{Podolsky04} of the specific problem of SDW coexisting with triplet SC, similar to that for the purely 1D systems. In another approach \cite{Grigoriev08} a semimetallic state of a non-fully gapped SDW is considered, where SC pairing occurs within electron and hole pockets.

Some other theories inferred spatial texture specifically for the SC and SDW coexistence and their reasoning is not applicable to metal/SDW separation at temperatures above the SC onset. In one of them the shape of SC domains is determined by the effective mass anisotropy \cite{Lee02}. Another theory infers square lattice of domains from the entanglement of SDW wave vector and the magnetic moment of triplet SC \cite{ZhangDeMelo06}.

An alternative scenario is based on the SDW order parameter variation in soliton phase (SP)\cite{GorkovGrigoriev05,GorkovGrigoriev07,Grigoriev09}. The SP theory suggests that SDW order parameter can become nonuniform with metallic(SC) domain walls emerging as sheets normal to the most conducting chain direction $\mathbf{a}$. However, recent transport anisotropy measurements in PF6\cite{Kang10} revealed that SC develops in a different manner, first in a form of filaments along the least conducting (interlayer) $\mathbf{c}$ direction and only for higher pressures transforms to 2D sheets.  Whereas this observation can resolve the inconsistency between the transport and NMR data, the origin of the spatial segregation and its anisotropy still remains unknown.

The emergence of the metallic(SC) phase along with SDW suppression is intimately related to the creation of ungapped carriers. The latter is governed by the Fermi surface (FS) nesting, which  easily occurs in  Q1D systems, where FS is a pair of slightly warped sheets. In experiments with PF6 the FS nesting is altered by pressure, which also affects lattice parameters. Indeed, under pressure nesting is spoiled due to FS warping caused by transfer integrals increase \cite{Yoshino01,Yamaji82}. This entanglement prevents one from separating the influence of variations of unnested carriers density from that of the lattice spacing changes with pressure. Consequently, to address this problem it is necessary to study the SC and SDW coexistence at a fixed pressure by changing some other external parameter controlling the FS nesting.

Such an opportunity is provided in (TMTSF)$_2$ClO$_4$ (hereafter denoted as ClO4), where the SC and SDW phases are known to coexist at ambient pressure \cite{Schwenk84, Gerasimenko13}. In this material, instead of pressure one can vary the degree of dipole ordering of ClO$_4$ tetrahedral anions. Due to slow kinetics of anion ordering\cite{Pouget90}, the latter can be varied over wide range by choosing the appropriate cooling rate in the vicinity of ordering transition temperature $T_{AO}=24$\,K. This possibility opens a way of exploring the remarkable phase diagram with a continuous transition from the SDW state at high disorders to the homogeneous SC state at low disorders \cite{Schwenk84,Gerasimenko13}.

Although the effect of anion ordering on conduction electrons is complicated, the most pronounced is the bandstructure folding in the interchain direction caused by the doubling of lattice period along the $\mathbf{b}$-axis in the ordered state. As a result, in the ordered state due to folding the FS splits in four sheets and nesting is spoiled. When anions are disordered the FS is not different from that in PF6 and is almost perfectly nested. It was suggested therefore, that SC/SDW coexistence in this system is due to the loss of long-range anion order as cooling rate is increased\cite{Schwenk84}.

An alternative explanation\cite{ZanchiBjelis01} suggests that the degree of ordering affects the magnitude of the FS splitting, the so-called dimerization gap $V$. It was shown theoretically\cite{ZanchiBjelis01,SenguptaDupuis01}, that the SDW onset temperature gradually decreases as the gap $V$ grows. This suggestion is supported by our recent angular magnetoresistance experiments\cite{Gerasimenko13}, where we traced the FS evolution with disorder. Particularly, upon increasing cooling rate, we observed gradual growth of interchain bandwidth (and hence decrease of the gap $V$) ending eventually with SDW onset. Furthermore, in the two-phase SC/SDW state the temperature dependences of resistance showed hysteresis\cite{Gerasimenko13}, which could not be ascribed to the loss of anion order. These observations allowed us to conclude, that the anion ordering induced FS splitting controls the phase coexistence.

The above results motivated us to perform measurements of the spatial texture of the SC phase embedded in the SDW background in ClO4 when only a single band parameter, the dimerization gap $V$, was varied. By tuning $V$ we drove the system through the SC/SDW phase boundary at the $T-V$ phase diagram, keeping the dispersion in the interlayer direction unaffected. We found that the SC regions arise first as filaments along $\mathbf{c}$ axis and then merge into sheets in the $\mathbf{b-c}$ plane. Our observations reproduce qualitatively the behavior observed in PF6 under pressure\cite{Kang10}. We conclude, therefore, that the SC spatial texture depends neither on details of band structure, as it was proposed in Ref.~\onlinecite{Kang10}, nor on character of lattice deformations under pressure. Such texture represents then the intrinsic way of allocating unnested carriers that appear with suppression of SDW state irrespective of their exact origin.

The observed anisotropy can be explained qualitatively within the soliton phase scenario for strongly anisotropic lattice, with no additional assumptions. Indeed, within the SP theory the energy gain comes from alignment of solitons in walls across the molecular chains, i.e. the creation of soliton band. The energy cost is primarily caused by the growing repulsion of solitons on the same chain as the number of walls increases\cite{BrazovskiiGorkovLebed82}. When one constructs a wall by adding another soliton in it, the width of the resulting soliton band will be much bigger for $\mathbf{b}$-axis due to larger transfer integral compared to that for $\mathbf{c}$-axis. This allows one to create longer walls along the $\mathbf{c}$-axis before the soliton band becomes larger than the SDW gap. This process leads to stronger repulsion (i.e., energy cost) of walls oriented along $\mathbf{b}$-axis, thus making  the $\mathbf{c}$-axis alignment of solitons more favorable. As the number of unnested carriers increases, the $\mathbf{c}$-axis soliton band will become fully occupied; further, the solitons will start aligning also along the $\mathbf{b}$-axis thus forming the $\mathbf{b-c}$-plane domains.

\section{Experimental}
The major experimental idea of our paper is to probe the SC/SDW phase diagram using the controlled variation of the degree of anion ordering. In this way we want to determine how the zero-resistance SC state emerges within the insulating SDW state in real space. Thus, the proper starting point for such experiment is to prepare our samples in the most homogeneous SDW state, i.e. to prevent anions from ordering as much as possible. This is achieved by rapid  cooling, or ``quenching'', of samples from some temperature $T_Q>T_{AO}$ above anion ordering temperature. Smaller degrees of ordering can be achieved by slower cooling, however such procedure would require careful quantification of the resulting disorders. Instead, we have chosen to anneal the quenched sample at some temperature below $T_{AO}$, which is also known to cause gradual ordering of anions\cite{Pouget90}. In this way we know exactly that each consecutive disorder is weaker. Furthermore, the use of annealing prevents us from affecting other sample properties except the degree of ordering after the initial disorder was created. In contrast, consecutive rapid coolings could result in uncontrollable change of sample state (e.g. create defects).

It should be emphasized though that quenched samples are essentially spatially inhomogeneous on a small scale. A prerequisite for this is the slow anion ordering \cite{Pouget90}, much slower than cooling rates used in experiment. As a result, small anion-ordered inclusions are formed in a disordered background. Whereas we cannot avoid creating them during cooldown, their number can be adjusted by the sample preparation. In what follows we present the results for various quench parameters and show the insignificant influence of these inclusions on anisotropy of the spatial texture of the SC phase.

In our experiment we vary the two independent parameters: quench rate and quench temperature. The former determine predominantly the degree of ordering, since the anion-ordered volume fraction tends to saturate for higher rates\cite{Pouget90}. The role of $T_Q$ is more complicated. On the one hand, in Ref.~\onlinecite{Schwenk84} the degree of ordering was found to increase as the quench temperature decreased. On the other hand, varying $T_Q$ changes fraction of the ordered inclusions  due to fluctuations of anion ordering noticeable below $\sim 30$\,K as observed in Ref.~\onlinecite{Zhang05}.

To characterize the resulting disorder we measured two parameters: the transition temperature and the width of the SDW transition. The former depends on the ordering-induced band-splitting $V$, namely, SDW onset at high temperatures reflects low degree of ordering \cite{Gerasimenko13,ZanchiBjelis01}. Typical values of transition temperatures achieved after quenching in our measurements, $6-6.5$\,K, are close to those observed experimentally by other groups \cite{Schwenk84,Qualls00}. The width of SDW transition is small for high quench rates and $T_Q$-temperatures; it increases as $T_Q$ decreases. Therefore, narrow and high-temperature SDW onset is the signature of weakly ordered state with low amount of anion-ordered inclusions. Wide and high-temperature onset indicates weakly ordered state with high amount of inclusions.

\begin{figure}[ht]
\begin{center}
\includegraphics[width=0.47\textwidth]{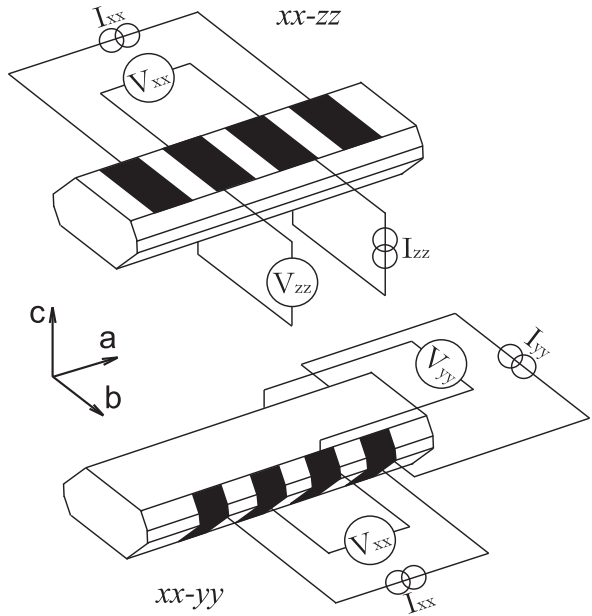}
\caption{Two resistivity measurement configurations used in this experiment: $xx-zz$ (top) and $xx-yy$ (bottom). Black rectangles show schematically contact pads (or graphite paint droplets) to which the Pt wires were glued.}
\label{fig_sample}
\end{center}
\end{figure}

We deduce the spatial structure of the two-phase state from resistance measurements. In our experiment we prepare samples in such a way, that two resistivity components can be measured on the same sample (see Fig.~\ref{fig_sample}). In this way we extract the anisotropy for the two of three planes by performing simultaneous measurements on different contacts for the same disorders. In the case of the third plane we quench two samples simultaneously to achieve close initial conditions and compare the results. To provide more support for these results we also measure SC critical fields, $H_{c2}$, and deduce from these data additional information on the spatial texture\cite{Lee02}.

(TMTSF)$_2$ClO$_4$ single crystals of typical dimensions $x\times y\times z=3\times0.1\times0.03$\,mm were synthesized by a conventional electrochemical technique. Three resistance components, $R_{xx}$, $R_{yy}$ and $R_{zz}$, along the orthorhombic principal axes, $\mathbf{a}$, $\mathbf{b}^\prime$ and $\mathbf{c}^*$, were measured  using conventional 4-wire low-frequency  AC-technique. In order to measure $R_{xx}$ and $R_{zz}$ ($R_{xx}$ and $R_{yy}$) components eight annealed 10\,um Pt wires were glued with a conducting graphite paint on the opposing faces normal to the $\mathbf{c}^*$ ($\mathbf{b}^\prime$) axis, correspondingly (see Fig.~\ref{fig_sample}). To relieve stress that could appear during cooldown, samples were left suspended above the holder on the long Pt wires. Since the crystals are small, measurements in $xx-zz$ and $xx-yy$ configurations were performed on separate samples, in order to avoid conduction shunting with contact pads. Each of two resistivity components on a single sample were measured sequentially. Care was taken to align contact pads opposite to each other to minimize the admixture of other resistivity components.

Resistances measured in this way do not represent precisely the corresponding bulk resistivities, because of somewhat inhomogeneous current distribution in the sample bulk. The latter affects mostly the $R_{xx}$ resistance in the $xx-zz$ sample. Indeed, in that case due to high resistivity anisotropy current spreads into the bulk only partially, which leads to the admixture of $\rho_{zz}$ component \cite{Buravov94}. In contrast, for $xx-yy$ sample $\rho_{zz}$ is shunted by the contact pads (see Fig.~\ref{fig_sample}), which allows us to separate intrinsic features of $\rho_{xx}$ temperature dependence from $\rho_{zz}$ ones. As we will show below, the $\rho_{zz}$ potential admixture does not affect the results of our observations.

Samples were mounted on a top-loading $^3$He probe and aligned at room temperature under microscope to within a few degrees for  measurements in magnetic field $\mathbf{H\|c^*}$. To improve temperature homogeneity during cooldown, holders with samples were put inside the copper casing. The probe with the sample was initially cooled from room temperature to a certain quench temperature $T_Q \gtrsim T_{AO} =24$\,K. The cooling rate at this stage was chosen $0.2-0.3$\,K/min in order to avoid cracks. Further the samples were quenched from $T_Q$ by rapidly bringing the sample holder in contact with the 1\,K stage. The quench rates were determined from either readings of the thermometer installed on the copper casing or from comparison of the sample resistance with the equilibrium $R(T)$ dependence, both demonstrating similar results. The cooling rates obtained this way were as high as 600\,K/min in the vicinity of $T_{AO}$. Somewhat smaller quench rates, e.g. 100\,K/min, were achieved after rapid heating the cold copper casing above $T_{AO}$ by its subsequent free cooling. After the sample was quenched, weaker disorders were obtained by annealing. Annealing temperatures were gradually increased from 15 to 23\,K in our measurements, causing progressively weaker disorders; annealing below 15\,K had negligibly small effect on $T_{SDW}$. To achieve the fully anion-ordered state, samples were cooled from 40\,K to 20\,K for 12 hours.

Quench cooling could in principle produce different kinds of inhomogeneities or strains in the sample. To see how this affects our data, we have measured low-temperature $R_{xx}(T)$-dependences on the opposite surfaces and, in addition, $R_{zz}(T)$-ones using various opposing contact pairs (see Fig.~\ref{fig_sample}). Although the resistance values were slightly different, all the essential features of the corresponding dependences, including the SDW onset, were reproduced with high accuracy on different contact pairs. Since SDW transition temperature is defined by the band splitting $V$, we deduce that there are no large-scale spatial inhomogeneities of anion ordering. This result is also in accord with the previous direct measurements of $V$ from angular dependences of magnetoresistance \cite{Gerasimenko13}.

The potential strains might originate from different thermal expansion coefficients of the sample and conducting paint used for making contacts. From the Debye temperature estimates of $\theta_D=213$\,K\cite{Garoche82}, we conclude that for $T_Q\le40$\,K the thermal expansion of a sample is negligibly small, according to the Gr\"uneisen formula\cite{Ziman72}, and almost no additional strain is produced by quenching. Rapid cooling might also relieve the strain generated at higher temperatures. It is believed to create microcracks in a sample, which in their turn can create parallel conduction channels predominantly along the interlayer $\mathbf{c}$-axis\cite{Oh04}. Such channel was suggested to alter $R_{zz}(\mathbf{H\|b^\prime})$ dependence, leading to saturation of magnetoresistance in high fields\cite{Oh04}. However, in our samples we did not observe significant shunting in fields up to 13\,Tesla, indicating the insignificance or absence of microcracks.

Our studies  have been done with six single crystals from two batches; all of them demonstrated qualitatively similar behavior. Below we present the results for the samples cut from three different crystals. The data for each of the three crystals is shown in Figs.~\ref{fig_xxzz}--\ref{fig_zzyy} correspondingly. The curves within a set of disorders are labeled relatively to each other in such a way, that \#1 corresponds to the most ordered state, and the highest number corresponds to the quenched state.

\section{Results}
\subsection{Anisotropy in the $\mathbf{a-c}$ plane}
\begin{figure}[ht]
\begin{center}
\includegraphics[width=0.45\textwidth]{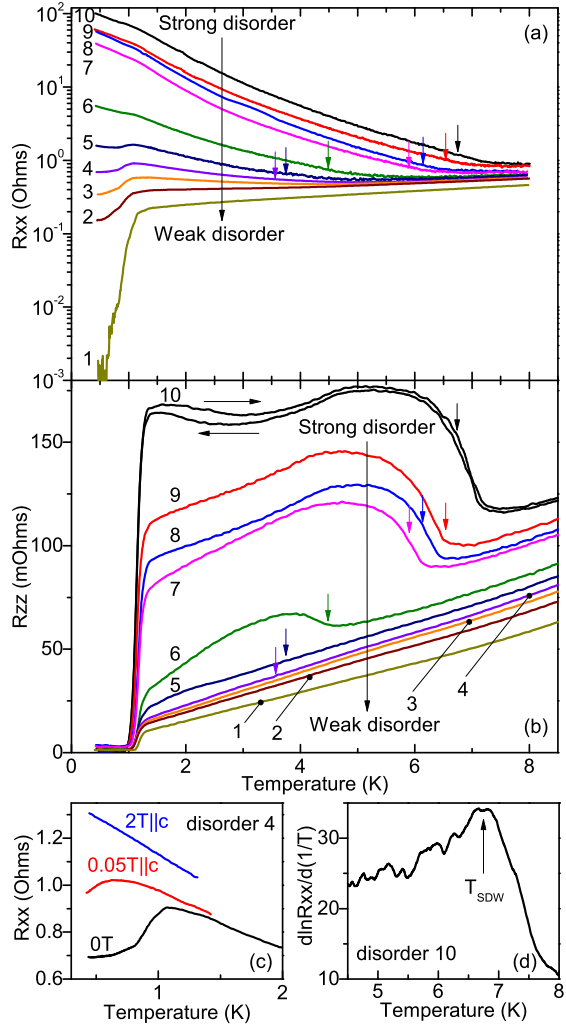}
\caption{Temperature dependences of (a) intralayer, $R_{xx}$, and (b) interlayer, $R_{zz}$, resistances for a set of disorders. Sample was cooled at the rate of approx. 100\,K/min from $T_Q=27$\,K. Arrows in panels (a) and (b) indicate SDW transition temperatures, $T_{SDW}$. The latter were determined as a peak value of $\mathrm{d}\ln R_{xx}/\mathrm{d}(1/T)$, as shown for the strongest disorder (\#10) in panel (d). Panel (c) shows magnetic field $H||c$ effect on $R_{xx}(T)$ dependence for one of disorders (\#4).
}
\label{fig_xxzz}
\end{center}
\end{figure}

The set of temperature dependences of $R_{xx}$ and $R_{zz}$ resistances for the same  sample at various disorders corresponding to different parts of the SC-SDW phase diagram is presented in Fig.~\ref{fig_xxzz}. This set covers the full phase diagram (Fig.~\ref{fig_xxzz}a) where the $R_{xx}(T)$ temperature dependence gradually changes its character from insulating to metallic and then to superconducting one. For strong disorders ($\#$6-10) the sample undergoes SDW transition, followed by insulating  temperature dependence characteristic of a gapped SDW state. The dependence is not activated, suggesting that the insulating background contains metallic inclusions. However, their fraction is apparently too small, below the percolation threshold, since no downturn in $R(T)$ (signalling the superconductivity onset) is observed at low temperatures, except for the kink at $T\approx1$\,K, where the $R_{xx}(T)$ slope becomes smaller. Remarkably, for the same disorders (\#\#6--10) $R_{zz}(T)$ demonstrates complete superconducting transition with $T_c \approx1$\,K  (see Fig.~\ref{fig_xxzz}\,b).  Moreover, the SDW onset seen in $R_{xx}(T)$ coincides with the steep increase of $R_{zz}(T)$ resistance at high temperatures, however  below $T_{SDW}$ it shows a downturn (\#\#6--10) in contrast to  $R_{xx}(T)$.

These observations show that at strong disorders the sample consists of metallic(SC) regions embedded in the SDW insulating background. Whereas they form a continuous SC path in the interlayer direction, inside the layers they are essentially separated from each other. The change of  the $R_{xx}(T)$ slope below 1K is observed for all the  measured samples and  is clearly connected with the SC onset. Indeed, the superconducting state emerging  inside the metallic regions is expected to improve shunting  of the SDW background at low temperatures due to emerging Josephson coupling between the neighboring regions.

The major effect of disorder in ClO4 is the variation of the band splitting $V$ - the deviation from perfect FS nesting, as was discussed above.  Therefore, as disorder weakens, the corresponding growth of the number of unnested carriers is expected to increase the metallic(SC) fraction; such behavior was indeed observed in PF6 upon approaching SDW endpoint\cite{Vuletic02}.

In accord with the latter expectation, one can see that superconductivity smoothly emerges on the  $R_{xx}(T)$-dependence  for weaker disorders (\#\#5--1), similar to the behavior observed in granular superconductors \cite{Deutscher80}. More detailed analysis indeed shows signatures of the percolation type transition. As disorder weakens, the dip appears at about 1K on the insulating $R_{xx}(T)$ dependence (curve \#5). Such reentrant behavior indicates the increase of metallic(SC) fraction, though it is still lower than percolation threshold. The dip is followed by metallic (\#4, \#3) temperature dependence, which suggests emergence of  superconducting paths  along the $\mathbf{a}$ axis. Indeed, both the downturn and the metallic behavior disappear in magnetic field (see Fig.~\ref{fig_xxzz}c), which unambiguously  indicates their superconducting nature.

As disorder  weakens, the superconductivity onset temperature remains almost constant, $T_c\sim 1$\,K, whereas the concomitant resistivity drop gradually enhances (\#2, \#3), until $R_{xx}$ vanishes to zero  along both axes, $\mathbf{a}$ and $\mathbf{c}$ (\#1). The latter is consistent with anion ordered ClO4 being a homogeneous superconductor \cite{Lebed08,Pesty88}. We note that such behavior of $T_c$ and $R(T)$ is similar to smooth development of the percolation transition in granular superconductors\cite{Deutscher80}.

\begin{figure}[ht]
\begin{center}
\includegraphics[width=0.45\textwidth]{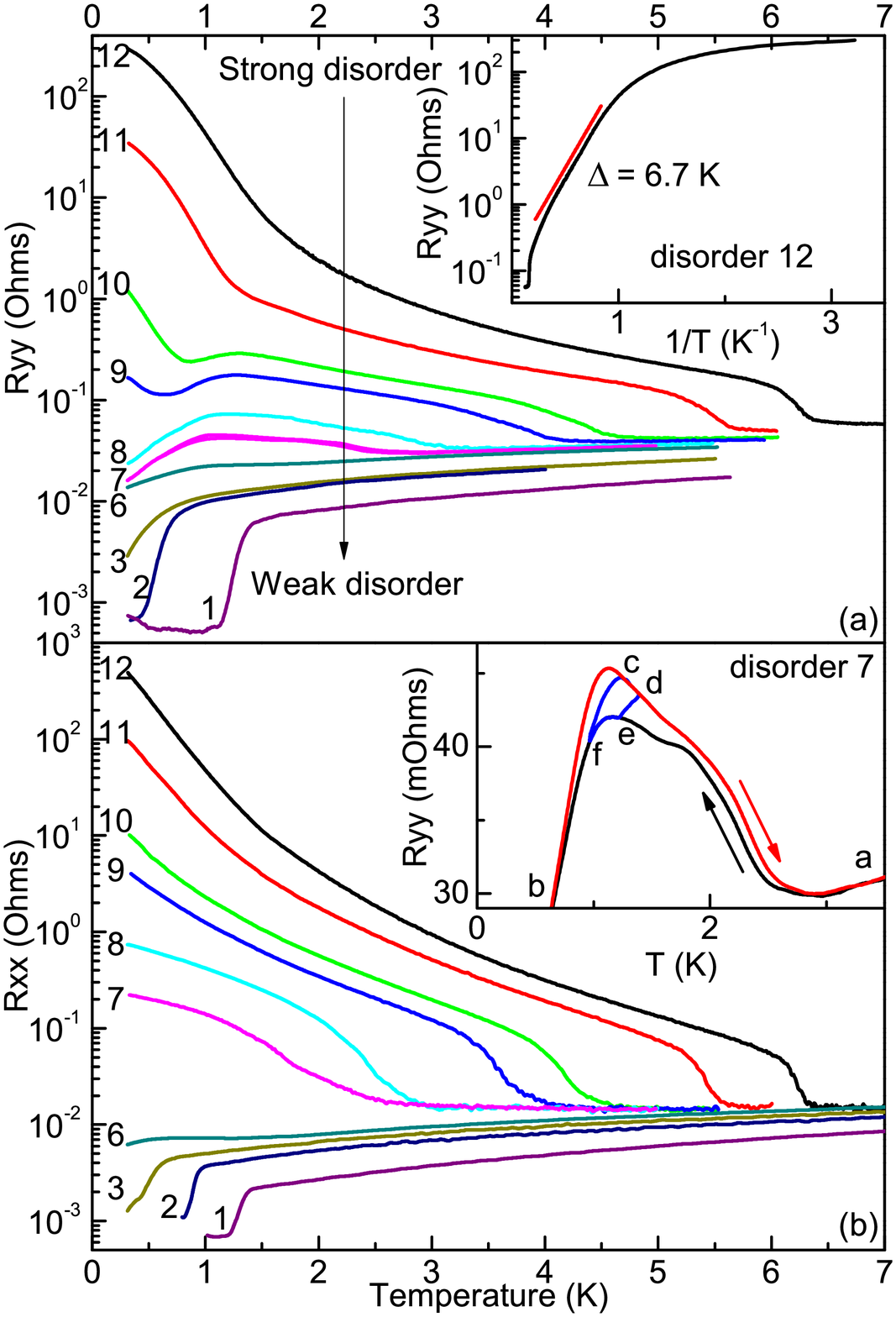}
\caption{Temperature dependences of intralayer (a) $R_{xx}$ and (b) $R_{yy}$ resistances for a set of disorders. Sample was cooled at the rate of approx. 600\,K/min from $T_Q=30$\,K. Inset in panel (a) shows the Arrhenius plot of $R_{yy}(T)$ for  the strongest disorder. Inset in panel (b) shows the hysteresis  $R_{yy}(T)$-dependence for disorder $\#7$. Two trajectories of temperature sweep: the large loop (a-e-f-b-c-d-a) includes $T_{SDW}\approx 2$\,K, whereas (c-d-e-f-c) takes place completely at $T<T_{SDW}$.}
\label{fig_xxyy}
\end{center}
\end{figure}

\subsection{Anisotropy in the $\mathbf{a-b}$ plane} Figure~\ref{fig_xxyy} shows disorder evolution of temperature dependences of the intralayer resistances $R_{xx}$ and $R_{yy}$. The SDW transition is much sharper for strong disorders and  the resistivity temperature dependence is closer to  the activated one, at least above $T\approx1$\,K (cf. inset on Fig.~\ref{fig_xxyy}a). Below 1\,K the slope of $R_{yy}(T)$  decreases; this decrease is lifted in small magnetic field of the order of the expected $H_{c2} \sim 0.1$\,T. We ascribe therefore the slowdown of $R_{yy}(T)$  to shunting caused by  small amount of metallic(SC) regions, same as discussed above for $xx-zz$ sample. This effect is more pronounced  for the latter sample  (e.g. \#10 in Fig.~\ref{fig_xxzz}a) which indicates larger amount of the metallic(SC) regions. This difference is connected with the lower quench rate in the vicinity of $T_{AO}$ as will be discussed in the following sections.

For the strongest disorder, both resistivity components show qualitatively similar temperature dependences (curves \#12 in Figs.~\ref{fig_xxyy}\,a,\,b). However, for somewhat weaker disorders (\#\#10--7) the $R_{yy}(T)$-dependence is non-monotonic; there firstly a dip (\#10, \#9) and then a downturn (\#8,\#7) appears below 1\,K on the insulating background.
For these same disorders, $R_{xx}(T)$-dependence remains monotonic and insulating. The non-monotonic features observed in $R_{yy}(T)$ indicate the superconductivity onset, similar to that described above for $xx-zz$ sample. As disorder weakens further (\#\#6--3), the downturn appears also in $R_{xx}(T)$, and finally both resistivity components show complete superconducting transition (\#2, \#1), as expected in the anion-ordered case. We emphasize, that the change of slope below 1\,K and appearance of the downturn in $R_{xx}(T)$ does not depend on the measurement configuration, $xx-zz$ or $xx-yy$, and thus adequately represents the $\mathbf{a}$-axis spatial texture.

Therefore, the results shown in figures \ref{fig_xxzz} and \ref{fig_xxyy} demonstrate that the metallic or superconducting paths emerge normal to the chains $\mathbf{a}$ axis.

\subsection{Hysteresis behavior} One of the most remarkable features of transport properties in the region of metal(SC) and SDW coexistence is the hysteresis in temperature dependence of resistances between cooling and heating curves (cf. \#10 on Fig.~\ref{fig_xxzz}b and \#7 on the inset in Fig.~\ref{fig_xxyy}b). When a sample is heated from  the SC state, the resistance is larger compared to that observed on  its cooling  from the high-temperature metallic state. Furthermore, the hysteresis vanishes only in the vicinity of the SDW transition. The hysteresis is a stationary effect and does not depend on the temperature sweep rate.  Both its width $\delta T\equiv T_\uparrow(R) -T_\downarrow(R)$ and amplitude $\delta R/R(T=\rm const)$ increase as disorder weakens. The hysteresis is observed only when the $R(T)$-dependence is insulating (i.e. the SDW state manifests). When $R(T)$ is metallic the cooling and heating traces coincided within experimental uncertainties  (see, e.g. curves \#\#3--1 in Fig.~\ref{fig_xxyy}a). Such a behavior clearly indicates history effects in the spatial distribution of the two-phase state and will be discussed  in more details later.

\subsection{Quench rate effect} Although  the maximum $T_{SDW}$ value for a given disorder is almost the same for $xx-zz$ and $xx-yy$ samples, somewhat different $R_{xx}(T)$ behaviors impede their direct comparison. For example, the dip on  the $R_{xx}(T)$ dependences which is present on curve \#5 in Fig.~\ref{fig_xxzz}a for $xx-zz$ sample is absent on curve $\#$10 in Fig.~\ref{fig_xxyy} for $xx-yy$ one with similar $T_{SDW}$ value. This fact indicates that the former sample has larger metallic(SC) phase fraction, as discussed above for the $xx-zz$ sample. This prevents one from figuring out the anisotropy of the SC phase spatial texture in the $\mathbf{b-c}$-plane. The only  parameter  different for the two samples is the quench rate, which was lower for $xx-zz$ one. Since the amount of metallic phase is lower for higher quench rate, the most direct comparison of the differences in the SC onset can be obtained for samples quenched (i) rapidly and (ii) at the same rate.

\begin{figure}[ht]
\begin{center}
\includegraphics[width=0.45\textwidth]{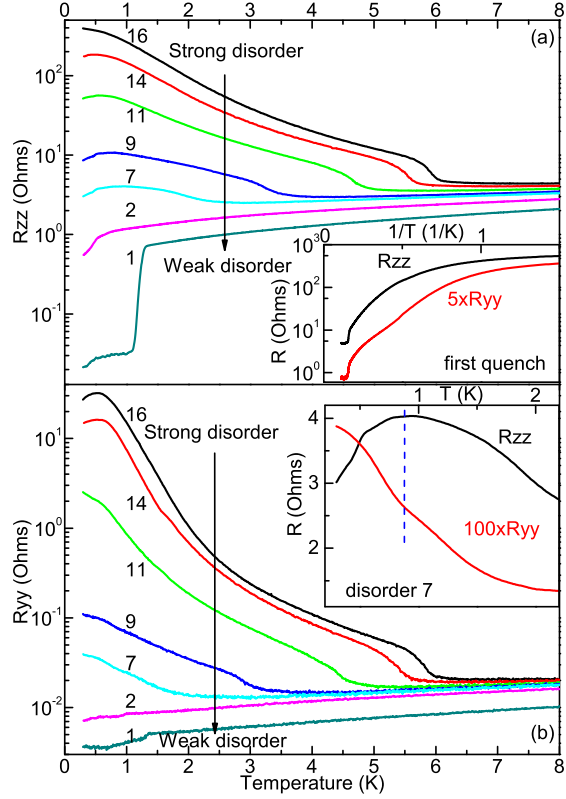}
\caption{Temperature dependences of (a) interlayer $R_{zz}$ and (b) intralayer $R_{yy}$ resistances for a set of disorders. Samples were cooled several times at the rate of approx. 600\,K/min from $T_Q=40$\,K. The downturn on $R_{yy}(T)$ for curves \#15, \#16 was absent for the first cooling (see inset in panel (a)) and appeared after numerous rapid coolings, and therefore might be an artifact. It is  also not pronounced at lower disorders. Inset in panel (a) shows the curves obtained after the first cooling of this sample at the rate of $\approx600$\,K/min. Inset in panel (b) shows the behavior of $R_{yy}$ and $R_{zz}$ on the same sample for moderate cooling rates. Dashed line indicates the position of the SC transition onset seen in $R_{zz}$ and the corresponding dip in $R_{yy}$.
}
\label{fig_zzyy}
\end{center}
\end{figure}

\subsection{Anisotropy in the $\mathbf{b-c}$ plane} To figure out the spatial anisotropy in the $\mathbf{b-c}$ plane we cut a sample in two parts to measure $R_{yy}$ and $R_{zz}$ separately and quenched both parts simultaneously several times at very high rates (up to 600\,K/min  in the vicinity of T$_{\rm AO}$) to achieve similar conditions. The results of the  $R_{yy}(T)$ and $R_{zz}(T)$ simultaneous measurements are presented in Fig.~\ref{fig_zzyy}. The abrupt SDW onset along with the pronounced activated temperature dependence indicate that the amount of the ordered phase is small,
as discussed above in relation with the quench rate effect. Already at this strongest disorder the $R_{yy}$ and $R_{zz}$ curves are notably different, though  the anion-ordered fraction is low and dimerization gap is small. Indeed whereas the $R_{yy}(T)$ dependence deviates  from activated behavior in the vicinity of the  SC onset, the $R_{zz}(T)$ dependence deviates at all temperatures indicating the substantial metallic shunting in the interlayer direction.

The curves at intermediate disorders show distinct difference between the $R_{yy}(T)$- and $R_{zz}(T)$-behaviors in the vicinity of SC onset (see inset in Fig.~\ref{fig_zzyy}b). The interlayer $R_{zz}(T)$ exhibits the downturn associated with the  SC onset (see also Fig.~\ref{fig_xxzz}c);  in contrast  the intralayer $R_{yy}(T)$ retains the insulating character with only minor deflection at $T_c$. Similar deflection transforming to a dip  was observed also in $R_{yy}(T)$ for $xx-yy$ sample (cf. \#11-9 in Fig.~\ref{fig_xxyy}a) and indicates the emergence of continuous SC paths along $\mathbf{b}$-axis.  Thus, the above measurements demonstrate that SC regions arise first along the $\mathbf{c}$-axis.

\begin{figure}[hb]
\begin{center}
\includegraphics[width=0.45\textwidth]{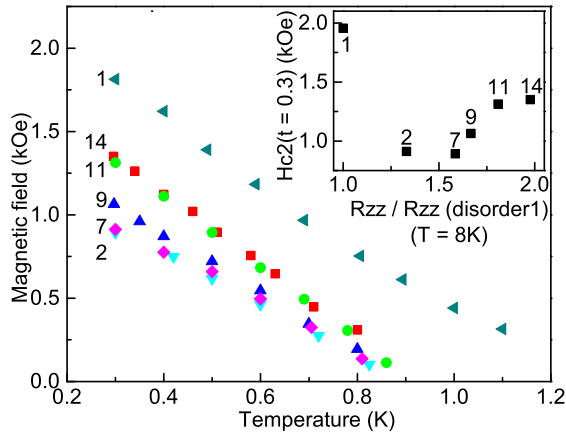}
\caption{Temperature dependences of critical field $H_{c2}||c^*$ for a set of disorders from Fig.~\ref{fig_zzyy}. $H_{c2}$ is determined as the onset ($99\%$) of resistive $R_{zz}(H)$ transition. Inset shows disorder dependence of critical field $H_{c2}$ at normalized temperature $t=T/T_c=0.3$. Various disorders on the horizontal axis  are characterized by the corresponding resistance $R_{zz}(T=8\,K)$ normalized to that in  the anion-ordered state.
}
\label{critfields}
\end{center}
\end{figure}

\subsection{Superconducting critical field} An independent probe for the in-plane texture of  a superconducting phase is the critical field $H_{c2}||c^*$. Both the critical field and the slope $dH_{c2}/dT$ for this direction are inversely proportional to  the in-plane coherence lengths $\xi_x \xi_y$. In the limit of small-sized SC regions, $d_i\le\xi_i$, the size of the region $d_i$ can be substituted for the coherence length $\xi_i$. Such a substitution can be justified at least for strong disorders \cite{note1}. Since we have figured out above that the SC paths emerge first  across the $\mathbf{a}$-axis, we conclude that $d_y$ grows much faster than $d_x$ with decreasing disorder. Therefore the growth of the critical field with disorder (at strong disorders) can be   associated with the domain size,  predominantly along the $\mathbf{b}$-axis, $d_y$.

There is no complete SC transition observed for strong disorders and the SC onset manifests only in the  $R_{zz}(T)$ downturn (seen, e.g. in Fig.3a). We use the disappearance of this downturn in magnetic field as an indicator for SC suppression and extract $H_{c2}$ in this way. The $H_{c2}(T)$-dependences are presented on Fig.~\ref{critfields}. One can see that both the slope $dH_{c2}/dT$ and $H_{c2}$ at low temperatures are approximately constant for strong disorders \#\#11--14 (see the inset in Fig.~\ref{critfields}). These observations demonstrate that for strong disorders \#\#11--14 the SC regions do not grow significantly in the $\mathbf{a-b}$-plane. In contrast, the enlargement of the regions along $\mathbf{c}$-axis is observed, as evidenced by the downturn that appears on insulating $R_{zz}(T)$-dependence below 1\,K for disorder \#14 but is absent for stronger ones (cf. \#16 on Fig.~\ref{fig_zzyy}).

For weaker disorders \#\#9--7 critical field decreases, indicating increase of $d_y$. This is also supported by the development of the deflection in $R_{yy}(T)$-dependence in the vicinity of $T_c$ (cf. inset in Fig.~\ref{fig_zzyy}b). Upon further decrease of disorder the low-temperature $H_{c2}$ tends to saturate (disorders \#\#7--2 in  the inset in  Fig.~\ref{critfields}), which is expected for $d_y>\xi_y$. The overall above behavior demonstrates, that as disorder weakens  the superconducting regions  grow first in the interlayer direction, and further, at intermediate disorders, start increasing along $\mathbf{b}$-axis.

Interestingly, $H_{c2}$  sharply increases again in the anion ordered state (see \#1 in Fig.~\ref{critfields}). Obviously, this effect has nothing to do with the spatial dimensions of SC regions, because the ordered state is believed to be the homogeneous superconductor\cite{Pesty88}. Similar behavior was also observed for weak disorders in (TMTSF)$_2$ClO$_4$\cite{Joo05} and (TMTSF)$_2$(ClO$_4$)$_{1-x}$(ReO$_4$)$_x$\cite{Joo04}. This intriguing issue however is out of the scope of the present paper and obviously requires further investigation.

\section{Discussion}
From our measurements of resistivity components and critical fields we conclude that (i) the spatial texture of the metallic(SC) phase emerging  in the background SDW state evolves from strongly anisotropic to almost isotropic, and (ii) its relative fraction   depends on prehistory. In this section we will discuss both  these results within two competing approaches. One of them is the emergence of soliton phase \cite{BrazovskiiGorkovLebed82,GorkovGrigoriev05}, which has intrinsic spatial anisotropy, and another is based on phase segregation  \cite{Vuletic02,Lee05} in anisotropic lattice. Firstly  we will address the question of whether  the existence of anion-ordered (AO) regions can be responsible for the observed anisotropy.

\subsection{Role of the AO}
Having no our own quantitative data on the fraction of the anion-ordered  regions  (where the metal(SC) phase is favorable), we can however make an estimate based on other available results. Earlier X-ray diffraction measurements\cite{Pouget90}  showed, that at a 100\,K/min quench rate, about 20\% of anion-ordered phase survives in small regions with average dimensions of $l_x=30$, $l_y=60$ and $l_z=50$\,nm along $\mathbf{a}$, $\mathbf{b}$ and $\mathbf{c}$ axes,  respectively. As disorder weakens, the number of AO inclusions grows while their size remains almost the same. Assuming  that the anion ordered inclusions have similar $l_y$ and $l_z$  size, we conclude that  this 20\% fraction is well below the 2D  percolation threshold in the $\mathbf{b-c}$-plane. Furthermore,  were  the superconductivity onset  a percolation transition in the $\mathbf{b-c}$-plane, it should be almost isotropic, which contradicts  the present results. Finally, such transition obviously cannot account for the hysteresis of the $R(T)$ dependence.

At the same time,  a larger number of AO inclusions, obtained at lower quench rates, apparently increases the  number of metallic(SC) regions without affecting the  whole spatial texture. Indeed,  in this case  we observe a  drastic difference in $R_{zz}(T)$ behavior: for rapidly quenched sample ($\sim600$\,K/min in the vicinity of $T_{AO}$) the transport along $\mathbf{c}$-axis is insulating (e.g. \#16 on Fig.~\ref{fig_zzyy}a), whereas at lower quench rates ($\sim100$\,K/min) it shows  signatures of a metallic behavior (e.g. \#9 on Fig.~\ref{fig_xxzz}\,b). Our simple explanation is that  the AO inclusions may  bridge the emerging metallic(SC) regions; the former are almost isotropic \cite{Pouget90} and  do not affect the anisotropy of the overall spatial texture.

\subsection{SDW suppression by imperfect nesting} Since anion ordering can account for neither the anisotropy, nor hysteresis, we conclude that the observed two-phase state is intimately related to SDW properties. The SDW is being suppressed along with the creation of unnested carriers. When pressure is a driving parameter (as in PF$_6$), nesting is most strongly deteriorated by 2D dispersion, essentially the increase of interchain ($b$-axis) hopping. Within the tight-binding model the nesting deviations are represented predominantly by the next-to-nearest interchain transfer integral, $t_b^\prime$\cite{Yamaji82}.

On the other hand, for Q1D systems it is known that magnetic field $\mathbf{H\|c}$ tends to reduce the size of electrons wave function to a single chain (see e.g. \cite{Lebed08}), thus acting opposite to pressure. This effect was predicted to restore SDW onset at higher temperatures\cite{Montambaux88}. The predictions of this so-called imperfect nesting theory were indeed observed in various $T_{SDW}(H,P)$ measurements \cite{Danner96,Gerasimenko09,Biskup95,Matsunaga01}. In the same way, measurements of $T_{SDW}$ behavior in magnetic field $\mathbf{H\|b}$ allows one to investigate the effect of interlayer dispersion. The latter due to its small value was observed to play minor role in SDW suppression \cite{Danner96,Gerasimenko09}.

The above effect of pressure is observed for both PF6 and ClO4\cite{Kang93}, but for the latter compound nesting can be spoiled also by increasing the band splitting  parameter $V$. It originates from charge disproportion between the neighboring chains in the $\mathbf{a-b}$ plane caused by displacement of ClO$_4^-$ tetrahedra \cite{Pouget90,Zhang05,LePevelen01} with respect to  the TMTSF molecules. Thus, the effect of the band splitting is quite similar to that of doping, but instead of changing number of carriers in a fixed bandstructure, the anion ordering shifts bands with respect to a fixed chemical potential. This is reminiscent of ``physical doping'' in layered  cuprates  where  under applied  electric field electrons are transferred  between neighboring layers \cite{Koval10}.

As the deviations from nesting increase, the indirect gap $\Delta_0-\epsilon(\mathbf{k}_\perp)$ may become negative, e.g. the bottom of conduction band in the center of Brillouin zone may become lower than the top of valence band at its border \cite{BrazovskiiGorkovSchrieffer82}. As discussed in Refs.~\onlinecite{BrazovskiiGorkovSchrieffer82,BrazovskiiGorkovLebed82}, at this point either electron and hole  pockets will appear atop  of the gapped Fermi surface, or the excess carriers can be trapped into solitonic midgap states resulting in  a non-uniform SDW order parameter. The former scenario can naturally lead to the phase separation if the SDW energy depends on deformation\cite{Lee05}, whereas the latter one incorporates a built-in spatial texture coming from alignment of solitons.

\subsection{Soliton phase scenario}
The energy cost  of a single soliton creation in a molecular chain is fixed and depends only on the SDW direct gap, $2\Delta_0/\pi$. The alignment of individual solitons in a wall across the chains  allows  double occupation of soliton levels and gains kinetic energy. The former   makes the  soliton band  on average half-filled \cite{BrazovskiiGorkovLebed82}. Kinetic energy is obviously larger for longer walls and for stronger hopping within  them.  Therefore, the soliton phase becomes favorable, if the density of unnested carriers, i.e. the length of a wall becomes high enough, so that the energy gain overcomes the  cost of creating solitons\cite{BrazovskiiGorkovSchrieffer82,BrazovskiiGorkovLebed82}. This process was  predicted to happen in PF6 before the semimetallic state emerges\cite{GorkovGrigoriev05}.

In three-dimensional case, the  soliton walls can be created along either $\mathbf{b}$ or $\mathbf{c}$-axis, or as 2D sheets in a $\mathbf{b-c}$-plane. The energy gain is the largest for 2D sheets, whereas 1D walls along $\mathbf{c}$-axis are the least favorable, because of small hopping integral in this direction. The above reasoning clearly contradicts spatial anisotropy observed in this paper.

The case of ClO4 is especially interesting in this context, since the emerging anisotropy is not the effect of the interplay of band parameters. The only quantity that grows as band splitting $V$ is increased is the number of unnested carriers. We observe that metallic(SC) domains are formed first along $\mathbf{c}$-axis and then merge/grow in the $\mathbf{b-c}$-plane sheets. This leads therefore to the quite clear scenario: with the increase of $V$ most of the unnested carriers are first allocated in the band of $\mathbf{c}$-axis quasi-1D filaments, and further they merge in 2D sheets in the $\mathbf{b-c}$-plane.

The major question following from the above discussion  is why the soliton walls initially emerge along the $\mathbf{c}$-  rather than the $\mathbf{b}$-axis? To answer this question, an interaction of solitons inside the molecular chains should be compared for these directions.
The energy losses increase strongly (at first, exponentially) as the inverse distance between the solitons within the same chain increases \cite{BrazovskiiGorkovLebed82}, i.e. the number of walls grows. In contrast, the energy gain increases linearly with a number of SWs. Therefore, creating longer walls is energetically favorable, though their length is limited by the maximum bandwidth, the SDW gap. At this point the difference between $\mathbf{b}$ and $\mathbf{c}$-axis walls comes in play. For a 1D filament the density of states increases with the distance between solitons within the wall, and thus is larger for $\mathbf{c}$-axis band due to the larger lattice spacing. Therefore, for the same width of soliton band the number of solitons within the $\mathbf{c}$-axis wall will be larger compared to that for $\mathbf{b}$-axis. This implies the larger number of walls in the latter case and hence the stronger interaction, which makes the alignment in $\mathbf{b}$ direction less favorable.  We note here, that SWs are not necessary, in theory, to be strictly 1D, but their length along $\mathbf{c}$-axis should be still much larger than that along $\mathbf{b}$-axis.

The described situation is somewhat different from that considered in Ref.~\onlinecite{BrazovskiiGorkovLebed82}, where the SP energy gain was calculated for the limit of small soliton bandwidth and, as a result, small number of walls. In experiment we observe the spatial anisotropy of metallic(SC) phase when SDW is nearly suppressed and $\epsilon$ is large\cite{Qualls00}. Therefore, both the soliton bandwidth and the number of SWs are implied to be large, and interaction between walls should be taken into account.

We would like to emphasize, that the above conclusion that the $\mathbf{c}$-axis is a preferable  direction, is inherent to the strongly-anisotropic systems with $t_b\gg t_c$, and should also apply to (TMTSF)$_2$PF$_6$ and other Bechgaard salts.

\subsection{Phase segregation scenario} One of the ways to allocate the unnested carriers is to create nonuniform deformation of the lattice, which (in the case of constant volume) will result in spatially separated metallic (compressed) and SDW (expanded) regions. The spatial dimensions of the regions in this case are defined by the interplay between the elastic and SDW energies. The direct way to explain the spatial texture is then to take into account the anisotropy of the lattice.

The major reason for this segregation to be favorable was pointed out by Lee et al.\cite{Lee05}: if the SDW energy gain depends on deformation $\delta x$, then the emergence of tricritical point and the first order transition follow from large
$\partial T_{SDW}/\partial x$ value in the vicinity of the SDW endpoint. In case of ClO4 we can track the evolution of spatial texture for fixed $b,c$ and $t_b,t_c$, as we get closer to the SDW endpoint and  increase $\partial T_{SDW}/\partial x$ by changing $V$.

The uniaxial lattice deformation along the $\mathbf{c}$-axis has the least energy cost, however the effect of hopping in this direction on nesting was shown to be small\cite{Danner96,Gerasimenko09}. Therefore, the spatial anisotropy within this scenario is not straightforward and more detailed  analysis is required. Below we utilize the model for phase segregation developed by Vuletic et al.\cite{Vuletic02} to estimate the fraction of metallic and SDW regions. To simplify the equations we treat the deformations along different axes independently, keeping in mind that due to Poisson ratio deforming e.g. $\mathbf{c}$-axis will result in much smaller effect on $\mathbf{a}$ and $\mathbf{b}$ axes. In this case the fraction of metallic phase along $i$-th axis is defined by\cite{Vuletic02}:
\begin{equation}
R_i=\frac{1}{2}-K_i|F_{SDW}(t_i)|\left(\frac{\partial F_{SDW}}{\partial t_i}\frac{\partial t_i}{\partial \delta_i}\right)^{-2},
\label{eq_psb}
\end{equation}
where $K_i$ and $\delta_i$ are elastic constant and the deformation along $i$-th axis and $F_{SDW}$ is the SDW energy gain. As $V$ increases, the only quantity affected in this relation is $|F_{SDW}(t_i,V)|(\partial F_{SDW}/\partial t_i)^{-2}$, whereas $K_i (\partial t_i/\partial \delta_i)^{-2}$ depends only on crystal structure and is therefore fixed. The former is smaller for $i=b$ as was shown above, whereas the latter is likely to be smaller for $i=c$. Indeed, $K_c<K_b$ and quasi-uniaxial stress experiments performed on PF6\cite{Guo00} showed large effect of $\mathbf{c}$-axis stress on interlayer transport, which indicates large value of $\partial t_c/\partial \delta_c$.

To account for the observed spatial anisotropy of the metallic phase we have to assume $R_c>R_b$ for strong disorders and $R_c\approx R_b$ for weaker disorders, close to the SDW endpoint. Qualitatively, for almost perfect nesting both $\partial F_{SDW}/\partial t_b$, $\partial F_{SDW}/\partial t_c$ are small, whereas for almost suppressed SDW the former is expected to be larger, reproducing the desired behavior. However, for this scenario to work,  one requires a certain relation between the elastic SDW properties  entering the Eq.~(\ref{eq_psb}), and is not a general property of an anisotropic system.

\subsection{Hysteresis} Hysteresis in ClO4 (and also in PF6\cite{Vuletic02}) is quite different from the one associated with the existence  of metastable phases at the first order transition. The latter occurs upon crossing the phase boundary by varying e.g. temperature. However, in our measurements the hysteresis is observed not only when the phase boundary is crossed, but also when the temperature is cycled well below $T_{SDW}$ (see inset in Fig.~\ref{fig_xxyy}b). These observations suggest that the hysteresis is linked rather with temperature dependent redistribution of phases.

Within the phase segregation scenario, such hysteresis can be ascribed to changes of the  spatial distribution of strain. Non-uniform lattice deformations are favorable as long as their cost is smaller than the overall energy gain, $F_\mathrm{metal}-F_\mathrm{SDW}\gtrsim K\delta_x^2$. In other words, the cost of local  strain variations should be less than that of metal-SDW transition. Thus, upon sweeping the temperature one can expect  transitions between almost degenerate local free energy minima  which correspond to  different strain distributions. The hysteresis then is caused by an effective ``dry friction'' associated with small energy barriers between the above minima. Within the SW scenario, the hysteresis might, in principle, originate from  pinning of the walls by defects, though no detailed treatment of the SP behavior in the vicinity of its boundary with metal was given up to date.

\subsection{Superconducting state}
Finally, we would like to comment briefly on the nature of the SC state coexisting with the SDW phase in ClO4. In this paper we use the  SC transition simply as a tool to determine the spatial texture  of the two-phase state, whereas the mere fact of the SC survival inside the disordered regions is intriguing itself. Indeed, it was suggested\cite{Yonezawa12} that in the anion-ordered state ClO4 is a nodal d-wave superconductor. Furthermore, random anion potential was shown to decrease $T_c$ at low disorders and therefore was suggested to act as a pair-breaking mechanism \cite{Pesty88,Joo05}. At higher disorders, where the two-phase state emerges, $T_c$ was noted to  saturate \cite{Joo05}; it was suggested therefore that the  SC phase is preserved in the anion-ordered inclusions\cite{Joo05} as SDW phase emerges inside disordered regions. The latter picture however contradicts our angular magnetoresistance measurements \cite{Gerasimenko13} and as was shown above cannot account for the anisotropy of the SC spatial texture.

It is also interesting to note, that somewhat similar saturation of $T_c$ with disorder was observed in quasi-2D superconductors, $\kappa$-(BEDT-TTF)$_2$X with X=Cu(NCS)$_2$\cite{Analytis06,Sano10} or Cu[N(CN)$_2$]Br\cite{Sano10}. The disorder in that case was created by either X-ray or proton irradiation. It is even more interesting that $T_c$ decrease in these compounds was also observed with varying cooling rate\cite{Su98}. The latter was suggested to freeze intrinsic disorder in BEDT-TTF molecule's methyl group orientations.

Despite these similarities, the underlying physics in these compounds is quite different from that in ClO4: the ground state competing with SC in them is Mott insulator. Therefore, the unifying property is likely the built-in molecular or anionic degrees of freedom. Indeed, these systems demonstrate explicitly the interaction between the conduction electrons and either molecular\cite{Kuwata11} or anion\cite{Zhang05} surrounding, as was deduced from NMR measurements. The peculiar nature of this interaction with additional degrees of freedom could be responsible then for the SC preservation even at high disorders. Obviously, more detailed studies are required to clarify the nature of superconductivity both with respect to disorder and inside the two-phase state.

\section{Conclusions}
In summary, we have studied the evolution of the spatial texture of metallic(SC) phase emerging inside the SDW phase as the latter is being suppressed by anion ordering in (TMTSF)$_2$ClO$_4$. The degree of ordering affects, on average, the band splitting $V$ responsible for SDW suppression\cite{ZanchiBjelis01,Gerasimenko13}, i.e. acts similar to doping. This is in contrast to the case of other TMTSF-based superconductors, where SDW vanishes due to increasing FS warping under pressure. We used controlled variation of anion disorder to disentangle the effect of unnested carriers from that of the lattice deformation and to determine the intrinsic spatial texture of the two-phase SC/SDW state.

We established that (i) metallic(SC) regions initially arise inside the majority SDW phase in a form of quasi-1D filaments elongated in the interlayer, $\mathbf{c}$-axis, direction; (ii) these regions merge in the sheets in the $\mathbf{b-c}$-plane as SDW is further suppressed by weakening disorder; (iii) although temperature dependence of resistivity exhibits hysteresis in the coexistence region, the spatial texture is independent of temperature sweep direction, i.e. prehistory. Furthermore, (iv) small anion-ordered regions embedded into the homogeneously disordered background increase the relative fraction of metallic(SC) phase (as evidenced  e.g. by the violation of activated behavior for strong disorders) but do not affect its spatial anisotropy. The similarity of anisotropy of the spatial texture in our measurements in (TMSTF)$_2$ClO$_4$ to that observed by Kang et al.\cite{Kang10} in (TMTSF)$_2$PF$_6$ suggests that the properties (i)-(iii) are intrinsic to metal(SC) and SDW coexistence in various organic quasi-1D systems.

We explained qualitatively the observed evolution of the spatial texture within the soliton phase theory\cite{BrazovskiiGorkovLebed82,GorkovGrigoriev05} in a strongly anisotropic lattice by taking into account the interaction between soliton walls. Still, the origin of hysteresis within this approach remains under question. The alternative phase segregation scenario turned out less promising, because in order to account for the anisotropy it requires some special assumptions on elastic properties of the lattice.

Recent generalization of soliton phase to quasi two-dimensional systems\cite{GorkovTeitelbaum10} could make our observations applicable also to the SC/SDW phase separation in iron pnictide superconductors.

\section{Acknowledgements}
The authors would like to thank P.D. Grigoriev, A.G. Lebed and S.E. Brown for discussions. The work was supported by RFBR, Programs of the Russian Academy of Sciences, by Russian Ministry for Education and Science (grant No 8375), and using research equipment of the Shared Facilities Center at LPI.


\begin{thebibliography}{99}
\bibitem{Lebed08}For review see: ``The Physics of Organic Conductors and Superconductors'', ed. A. Lebed, Springer-Verlag (2008)
\bibitem{Stewart11}G.R.~Stewart, Rev. Mod. Phys., {\bf 83}, 1589 (2011).
\bibitem{Johnston10} D.C.~Johnston, Adv. Phys., {\bf 59}(6),  803–1061 (2010).
\bibitem{Chubukov12} A.~Chubukov, Ann. Rev. Cond. Mat. Phys.  {\bf 3}, 57–92 (2012)
\bibitem{Vuletic02} T. Vuletic, P. Auban-Senzier, C. Pasquier, S. Tomic, D. J\'erome, M. H\'eritier, K. Bechgaard, Eur. Phys. J. B \textbf{25}, 319 (2002);
\bibitem{Kornilov04} A.V. Kornilov, V.M. Pudalov, Y. Kitaoka, K. Ishida, G.-q. Zheng, T. Mito, J. S. Qualls, Phys. Rev. B \textbf{69}, 224404 (2004);
\bibitem{Lee05}I.J. Lee, S.E. Brown, W. Yu, M.J. Naughton, P.M. Chaikin, Phys. Rev. Lett. \textbf{94}, 197001 (2005)
\bibitem{Yu02}W. Yu, S.E. Brown, F. Zamborszky, I.J. Lee, P.M. Chaikin, Int. J. Mod. Phys. B \textbf{16}, 3090 (2002)
\bibitem{Lee02} I.J. Lee, P.M. Chaikin, M.J. Naughton, Phys. Rev. Lett. \textbf{88}, 207002 (2002)
\bibitem{Podolsky04}D. Podolsky, E. Altman, T. Rostunov, E. Demler, Phys. Rev. Lett. \textbf{93}, 246402 (2004)
\bibitem{Grigoriev08}P.D. Grigoriev, Phys. Rev. B \textbf{77}, 224508 (2008)
\bibitem{ZhangDeMelo06}W. Zhang and C.A.R. S\'a de Melo, Phys. Rev. Lett. \textbf{97}, 047001 (2006)
\bibitem{GorkovGrigoriev05}L.P. Gor'kov and P.D. Grigoriev, Europhys. Lett. \textbf{71}, 425 (2005)
\bibitem{GorkovGrigoriev07}L.P. Gor'kov and P.D. Grigoriev, Phys. Rev. B \textbf{75}, 020507 (2007)
\bibitem{Grigoriev09}P.D. Grigoriev, Physica B \textbf{404}, 513 (2009)
\bibitem{Kang10}N. Kang, B. Salameh, P. Auban-Senzier, D. J\'erome, C.R. Pasquier, S. Brazovskii, Phys. Rev. B \textbf{81}, 100509 (2010)
\bibitem{Yoshino01}H. Yoshino, A. Oda, K. Murata, H. Nishikawa, K. Kikuchi, I. Ikemoto, Synth. Met. \textbf{120}, 885 (2001)
\bibitem{Yamaji82}K. Yamaji, J. Phys. Soc. Jap. \textbf{51}, 2787 (1982)
\bibitem{Schwenk84}H. Schwenk, K. Andres, F. Wudl, Phys. Rev. B, \textbf{27}, 5846 (1983); \textbf{29}, 500, (1984)
\bibitem{Gerasimenko13} Ya.A. Gerasimenko, V.A. Prudkoglyad, A.V. Kornilov, S.V. Sanduleanu, J.S. Qualls, V.M. Pudalov, JETP Lett. \textbf{97}, 419 (2013)
\bibitem{Pouget90} J.-P. Pouget, S. Kagoshima, T. Tamegai, Y. Nogami, K. Kubo, T. Nakajima, K. Bechgaard, J. Phys. Soc. Japan \textbf{59}, 2036 (1990)
\bibitem{ZanchiBjelis01}D. Zanchi and A. Bjelis, Europhys. Lett. \textbf{56}, 596 (2001)
\bibitem{SenguptaDupuis01}K. Sengupta and N. Dupuis, Phys. Rev. B \textbf{65}, 035108 (2001)
\bibitem{Buravov94} L.I. Buravov, N.D. Kushch, V.N. Laukhin, A.G. Khomenko, E.B. Yagubskii, M.V. Kartsovnik, A.E. Kovalev, L.P. Rozenberg, R.P. Shibaeva, M.A. Tanatar, V.S. Yefanov, V.V. Dyakin, V.A. Bondarenko, J. Phys. I France \textbf{4}, 441-451 (1994)
\bibitem{Garoche82} P. Garoche, R. Brusetti, D. J\'erome, J. Phys. Lett. \textbf{43}, L147 (1982)
\bibitem{Ziman72} J. M. Ziman, Principles of the Theory of Solids, Cambridge University Press (1972)
\bibitem{Oh04} J.I. Oh and M.J. Naughton, Phys. Rev. Lett. \textbf{92}, 067001 (2004)
\bibitem{Zhang05}F. Zhang, Y. Kurosaki, J. Shinagawa, B. Alavi, S.E. Brown, Phys. Rev. B \textbf{72}, 060501 (2005)
\bibitem{Pesty88}F. Pesty, K. Wang, P. Garoche, Synth. Met. \textbf{27}, 137 (1988)
\bibitem{Deutscher80}G. Deutscher, O. Entin-Wohlman, S. Fishman, Y. Shapira, Phys. Rev. B \textbf{21}, 5041 (1980)
\bibitem{note1} The $a$-axis BCS superconducting coherence length is $\xi_x=\hbar v_F/1.76 T_c\approx400$\,nm with $v_F=10^7$\,cm/s and $T_c=0.96\,K$. The charachteristic size of SC regions, $d_x$ for strong disorders can be estimated either from 30-60\,nm  sizes of AO inclusions or the width of the assumed soliton wall along $a$-axis. The latter is of the order of SDW coherence length $\xi_{SDW}=\hbar v_F/1.76 T_{SDW}\approx70$\,nm. Both these estimates for $d_x$ are smaller than $\xi_x$.
\bibitem{BrazovskiiGorkovSchrieffer82}S. A. Brazovskii, L. P. Gor'kov, J. S. Schrieffer, Phys. Scr. \textbf{25}, 423 (1982)
\bibitem{BrazovskiiGorkovLebed82}S. A. Brazovskii, L. P. Gor'kov, A. G. Lebed, JETP \textbf{56}, 683 (1982)
\bibitem{Montambaux88} G. Montambaux, Phys. Rev. B \textbf{38}, 4788 (1988)
\bibitem{Danner96}G.M. Danner, P.M. Chaikin and S.T. Hannahs, Phys. Rev. B \textbf{53}, 2727 (1996)
\bibitem{Gerasimenko09} Ya.A. Gerasimenko, V.A. Prudkoglyad, A.V. Kornilov, V.M. Pudalov, V.N. Zverev, A.-K. Klehe, J.S. Qualls, Phys. Rev. B \textbf{80}, 184417 (2009)
\bibitem{Biskup95} N. Bi\v{s}kup, S. Tomic, D. J\'erome, Phys. Rev. B \textbf{51}, 17972 (1995)
\bibitem{Matsunaga01}N. Matsunaga, K. Yamashita, H. Kotani, K. Nomura, T. Sasaki, T. Hanajiri, J. Yamada, S. Nakatsuji, H. Anzai, Phys. Rev. B \textbf{64}, 052405 (2001)
\bibitem{LePevelen01}D. Le P\'evelen, J. Gaultier, Y. Barrans, D. Chasseau, F. Castet, L. Ducasse, Eur. Phys. J. B \textbf{19}, 363 (2001)
\bibitem{Koval10} Y. Koval, X. Jin, C. Bergmann, Y. Simsek, L. \"{O}zy\"{u}zer, P. M\"{u}ller, H. Wang, G. Behr, B. B\"{u}chner, Appl. Phys. Lett. \textbf{96}, 082507 (2010)
\bibitem{Qualls00}J.S. Qualls, C.H. Mielke, J.S. Brooks, L.K. Montgomery, D.G. Rickel, N. Harrison, S.Y. Han, Phys. Rev. B \textbf{62}, 12680 (2000)
\bibitem{Kang93}W. Kang, S.T. Hannahs and P.M. Chaikin, Phys. Rev. Lett \textbf{70}, 3091 (1993)
\bibitem{Guo00} F. Guo, K. Murata, A. Oda, Y. Mizuno, H. Yoshino, J. Phys. Soc. Jap. \textbf{69}, 2164 (2000)
\bibitem{Joo05}N. Joo, P. Auban-Senzier, C.R. Pasquier, D. J\'erome, K. Bechgaard, Europhys. Lett., \textbf{72}, 645 (2005)
\bibitem{Joo04}N. Joo, P. Auban-Senzier, C. Pasquier, P. Monod, D. J\'erome, K. Bechgaard, Eur. Phys. J. B \textbf{40}, 43 (2004)
\bibitem{Yonezawa12}S. Yonezawa, Y. Maeno, K. Bechgaard and D. J\'{e}rome, Phys. Rev. B \textbf{85}, 140502(R) (2012)
\bibitem{Analytis06} J.G. Analytis, A. Ardavan, S.J. Blundell, R.L. Owen, E.F. Garman, C. Jeynes, B.J. Powell, Phys. Rev. Lett. \textbf{96}, 177002 (2006)
\bibitem{Sano10} K. Sano, T. Sasaki, N. Yoneyama, N. Kobayashi, Phys. B \textbf{405}, S279 (2010)
\bibitem{Su98} X. Su, F. Zuo, J.A. Schlueter, M.E. Kelly, J.M. Williams, Phys. Rev. B \textbf{57}, R14056 (1998)
\bibitem{Kuwata11} Y. Kuwata, M. Itaya, A. Kawamoto, Phys. Rev. B \textbf{83}, 144505 (2011)
\bibitem{GorkovTeitelbaum10} L.P. Gor'kov and G.B. Teitel'baum, Phys. Rev. B \textbf{82}, 020510 (2010)
\end{thebibliography}
\end{document}